# Electronic Band Structure of Wurtzite GaP Nanowires via Resonance Raman Spectroscopy

Jaya Kumar Panda[1], Anushree Roy[1*], Mauro Gemmi[2], Elena Husanu[3], Ang Li[3], Daniele Ercolani[3], and Lucia Sorba[3†]

[1]Department of Physics, Indian Institute of Technology Kharagpur. Pin 721302. India

[2]Center for Nanotechnology Innovation @ NEST, Istituto Italiano di Tecnologia, Piazza S. Silvestro 12, I-56127 Pisa, Italy

[3]NEST-Istituto Nanoscienze-CNR and Scuola Normale Superiore, Piazza S. Silvestro 12, I-56127 Pisa, Italy

Raman measurements are performed on defect-free wurzite GaP nanowires. Resonance Raman measurements are carried out over the excitation energy range between 2.19 and 2.71 eV. Resonances at 2.38 eV and 2.67 eV of the $E_1(LO)$ mode and at 2.67 eV of the $A_1(LO)$ are observed. The presence of these intensity resonances clearly demonstrates the existence of energy states with $\Gamma_{9hh}$ and $\Gamma_{7V}$ ($\Gamma_{7C}$) symmetries of the valence (conduction) band and allows to measure WZ phase GaP band energies at the $\Gamma$ point. In addition, we have investigated temperature dependent resonant Raman measurements, which allowed us to extrapolate the zero temperature values of $\Gamma$ point energies, along with the crystal field and spin-orbit splitting energies. Above results provide a feedback for refining available theoretical calculations to derive the correct wurtzite III-V semiconductor band structure.

---

[*] Email: anushree@phy.iitkgp.ernet.in
[†] Email: lucia.sorba@sns.it

## 1. Introduction

In recent times, novel crystal structures in III-V semiconductor nanowires (NWs) have generated immense scientific interest. The dimensions of these wires are too large to exhibit electron/phonon confinement [1-3]. However, the modified crystal structure in such wires (in reference to corresponding bulk samples) indicates new possibilities in the realm of nanoscience and nanotechnology [4-17] It is now well established that NWs of III-V semiconductors can be grown in wurtzite (WZ) phase along their length, though the corresponding bulk materials are in zinc-blende (ZB) phase [5]. The indirect band gap in ZB phase [18] limits the application of the latter in optoelectronic device fabrication. The recent prediction of a direct band gap and drastic changes in the electronic band structure and optical properties of III-V semiconductors [19, 20] n WZ phase motivated the researchers to investigate their physical properties through experiments.

The electronic band structure at $\Gamma$ point of WZ III-V semiconductor NWs have been probed using resonance Raman (RR) measurements. It is known that such measurements exploit the coupling of different phonon modes with the electronic states having different symmetry properties. The resonance of $E_1(LO)$ phonon mode corresponds to coupling of electronic energy states in the band structure with $\Gamma_{9hh} \rightarrow \Gamma_{7C}$, $\Gamma_7 \rightarrow \Gamma_{7C}$ and $\Gamma_{9hh} \rightarrow \Gamma_{8C}$ transitions. On the other hand, the allowed intensity resonance of the $A_1(LO)$ mode corresponds to $\Gamma_7 \rightarrow \Gamma_{7C}$ symmetry states of the electronic structure [10, 21] Here $\Gamma_7$ can be either $\Gamma_{7V}$ or $\Gamma_{7lh}$ depending on the electronic band structure of the particular system. Thus, one expects resonance of $E_1(LO)$ mode for three different excitation energies in the optical range (corresponding to three different coupling of electronic energy states of different symmetries), one of which should match with the resonance energy of the $A_1(LO)$ mode. Till date most of the works reported in this area are on arsenic based WZ III-V semiconductor NWs. From RR measurements of the 2LO mode of GaAs NWs, Ketterer et al [10]observed Raman resonance, coupling $\Gamma_{9hh} \rightarrow \Gamma_{7C}$ symmetry state of GaAs NWs. The possibility of coupling of $\Gamma_{7lh} \rightarrow \Gamma_{7C}$ states is excluded as 2LO scattering efficiency is negligible for this case. Peng et al [9]discussed the electronic band structure of GaAs NWs using RR measurements of first order LO mode. However, the expected resonances of $E_1$(LO) mode at different energies coupling different symmetry states are not reported. The incoming and outgoing resonances of LO line profile have been attributed to coupling $\Gamma_{9hh} \rightarrow \Gamma_{7C}$ and $\Gamma_{7lh} \rightarrow \Gamma_{7C}$ states in GaAs NWs. The paper is silent about the selection rule followed by above individual LO modes of WZ phase in electron-phonon coupling [7]. For AlAs NWs as the energy scale for the coupling of $\Gamma_{7lh} \rightarrow \Gamma_{7C}$ states is beyond the optical range, the resonance of $A_1$(TO) mode could not be observed and only low energy resonance peak is reported for the $E_1$(TO) mode [17]. Direct band gap of WZ GaP NW has recently been observed from photoluminescence measurements [22].

As for arsenic based semiconductors, bulk GaP is usually found in the ZB phase. In the present work, we grow defect-free WZ GaP NWs and study its structural properties by transmission electron microscopy (TEM. The electronic properties are investigated by RR measurements. It is to be noted that the energy scale of the electronic states at $\Gamma$ point with above mentioned symmetries for GaP NWs of WZ phase is within the range of our probing excitation wavelengths and the RR scattering is expected to identify these energy states of WZ GaP NWs without any ambiguity. In addition, we investigate the temperature variation of the energy gap of GaP NWs using temperature-dependent RR spectroscopy.

## 2. Experimental

The NWs investigated in this work were grown by Au assisted chemical beam epitaxy (CBE) technique in a Riber Compact 21 system on a (111) B GaAs wafer. The system employs pressure control of the metal organic (MO) precursors in the lines to vary their fluxes during the sample growth. Tertbutylarsine (TBAs), tertbutylphosphine (TBP) and triethylgallium (TEGa) were the MO precursors involved in the growth of GaP NWs with GaAs stem. Due to their higher decomposition temperature, TBAs and TBP were pre-cracked in the injector at 1000 °C. A 0.5 nm thick Au film was pre-deposited on the GaAs wafer by thermal evaporation in a separate chamber and then transferred to the CBE growth chamber. In order to remove the native oxide and to form Au nanoparticles by thermal dewetting of the Au film, the wafer was annealed at 560 °C for 20 minutes under TBAs flow. For the growth of GaAs stem TEGa flow was introduced in the chamber at a growth temperature of T=560 °C. To change from GaAs to GaP growth, TEGa flux and substrate temperature were kept constant while abruptly switching the V precursor flux.

Raman spectra were collected in back scattering geometry by using a micro-Raman setup equipped with $Ar^+$-$Kr^+$ laser (Model 2018-RM, Newport, USA) as the excitation light source, a spectrometer (model T64000, JY, France) and a Peltier cooled CCD detector. The backscattered signal was collected through a 100× objective for room temperature and 50×objective for low/high temperature measurements. The scattered light was dispersed by an 1800 $mm^{-1}$ grating under triple subtractive mode. The size of the laser spot at the sample surface was approximately 1 μm for 100× and 2 μm for 50×. For RR measurements, we employed several excitation lines between 457 nm and 568 nm from $Ar^+$-$Kr^+$ laser. A low incoming laser power was chosen to avoid the heating effect. Temperature dependent Raman measurements were carried out by using a sample cell (Model Link-600 S, Linkam Sci. Instr., UK). We allowed at least 30 minutes to stabilize the set temperature before the measurements. For polarized Raman scattering measurements, a λ/2 plate in the path of incoming beam (along x direction) was used, so that the incident polarization could be rotated. The scattered light was recorded by positioning an analyzer in front of the spectrometer slit with its pass axis either in y or z direction. Unless otherwise mentioned the Raman measurements were carried out on GaP NWs transferred on a Si substrate. The spot was focused on the GaP segment of the wire.

## 3. Results and Discussion

The crystal structure of the GaP NWs was investigated with a Zeiss Libra 120 TEM operating at an accelerating voltage of 120 kV. Crystal structure determination and defect analysis of the NWs was carried out by transferring the NWs to copper TEM grids coated with a formvar carbon film and analyzing both the TEM micrographs and the electron diffraction patterns. Figure 1(a) is a bright field image of a typical NW acquired in the $[2\bar{1}\bar{1}0]$ WZ zone axis. The GaAs stem on the right end of the NW, with darker contrast, shows a lot of defects and small segments of ZB, appearing as thin, sharp lines perpendicular to the growth direction. On the other end, the ~1μm-long GaP segment near the NW tip (identified by the dark AuGa alloy particle on the left end of the NW) appears completely free of crystal defects. Figure 1(b) shows the electron diffraction pattern taken on the GaP segment. The pattern can be indexed as $[2\bar{1}\bar{1}0]$ zone axis of a WZ crystal structure, and there are no sign of ZB diffraction spots nor the typical streaking parallel to (0001) reciprocal direction, which would indicate the presence of crystal defects.

Figure 2 shows a typical room temperature Raman spectrum of our WZ GaP NWs. The phonon dispersion for WZ phase can be approximated by folding phonon dispersion (Γ→L) for ZB (111) structure. As a result of the folding, new phonon modes are expected to appear in the Raman spectrum of the corresponding WZ phase [23]. Table I summarizes the expected phonon modes in WZ phase of GaP under different polarization configurations. Mode wavenumbers are obtained by folding the phonon dispersion curve of GaP of ZB phase [24] at mid point. Listed polarization configurations (following Porto's notation) are expected from the group theory analysis of similar III-V WZ systems [8, 23]. The *x,y,z* axes represent the crystallographic directions [11$\bar{2}$0], [1$\bar{1}$00] and [0001], respectively. Other than Raman active $E_1$(TO) and $A_1$(LO) modes at 361 and 402 cm$^{-1}$, folding of these modes at L point gives rise to two new phonon modes at 355 cm$^{-1}$ ($E_2^H$) and 375 cm$^{-1}$ ($B_1^H$). Folding of acoustic phonon modes result in extra modes at 215 cm$^{-1}$ ($B_1^L$) and 86 cm$^{-1}$ ($E_2^L$). $B_1^H$ and $B_1^L$ are expected to be Raman silent. The anisotropy of the atomic bonds along or in the perpendicular directions of the c-axis results in mixing of $A_1$ and $E_1$ modes. Hence two more phonon modes, $E_1$(LO) and $A_1$(TO), are expected to appear. It must be pointed out that the wavenumbers obtained with this simple folding of ZB phonon dispersion are to be considered as indicative, and discrepancies with the measured wavenumbers are to be expected. The correct theoretical estimation of the peak position needs detailed group theory analysis of GaP in WZ phase which goes well beyond the scope of the present work.

For the data of Fig. 2, the best fit could be obtained by deconvoluting the spectrum with five Lorentzian functions for the $A_1$(TO)/ $E_1$(TO) mode, LO modes and $E_2^H$ mode. We keep the intensity, peak position and width as free fitting parameters. Deconvoluted components are shown by dashed lines in Fig. 2. $E_1$(LO), $A_1$(LO), $A_1$(TO)/ $E_1$(TO), $E_2^H$ modes appear at 397, 391, 361 and 353 cm$^{-1}$, respectively. Though expected to be Raman silent, $B_1^H$ could be observed in the Raman spectrum of GaP NW at 383 cm$^{-1}$. This mode has been recently observed in WZ AlAs NWs and has been attributed to be due to the stacking faults of crystal structure along the length of the wire [17]. In our case, though, the main stem of the GaP NW is completely free of crystal defects (as seen in Fig. 1 (a)); we believe that GaAs/GaP interface results in breakdown of crystal symmetry at the junction and hence results in appearance of the silent $B_1^H$ mode. $E_2^L$ $x(zz)\bar{x}$ mode appears at 80 cm$^{-1}$. Inset of the Fig. 2 shows the Raman spectrum of single NW in polarization configuration with z along the axis of the wire. As expected the spectrum is dominated by allowed $A_1$ (TO) mode. Weak LO modes are also observed in the spectrum. The appearance of the forbidden LO modes may be due to slight deviation of the experimental set up from the true polarization configuration. Measured Raman mode wavenumbers are also tabulated in Table I.

RR scattering spectroscopy is a powerful noninvasive technique to map the electronic band structure at specific symmetry points of a system via electron-phonon interactions. We probe the band gap at Γ point of GaP NWs in WZ phase using RR profiles of $A_1$(LO) and $E_1$(LO) modes. Room temperature Raman spectra of GaP NWs recorded for different excitation energies over the range between 2.19 and 2.71 eV are shown in Fig. 3 (a). Each spectrum was fitted with five Lorentzian functions for five different Raman modes (as observed in Fig. 2). Variation of integral intensity of $A_1$(LO) and $E_1$(LO) modes with excitation energy are shown in Fig. 3 (b). The solid line is a guide to the eye. Intensity variation of the $E_1$(LO) mode show resonances at 2.38 and 2.67 eV, whereas, $A_1$(LO) exhibits resonance at 2.67 eV.

In the ZB phase, GaP is an indirect band gap semiconductor with conduction band minima at X, L and Γ points in sequence. The electronic band structure of WZ phase of GaP has been calculated using transferable empirical pseudopotential method including spin-orbit coupling and is shown to be very different from that of its ZB phase [19]. Spin-orbit and crystal field perturbation in WZ structure results in splitting of $\Gamma_{15}$ valence state of cubic ZB phase into $\Gamma_{9hh}$ heavy hole, $\Gamma_{7lh}$ light hole and $\Gamma_{7V}$ crystal field split off valence bands. Though, in general, for the WZ structure the lowest conduction band minimum has $\Gamma_7$ symmetry, above calculation predicts it to be of $\Gamma_8$ symmetry for few III-V semiconductors (GaAs, AlAs, InAs, GaP).

The schematic diagram of the electronic band structure near the $\Gamma$ point of the WZ GaP, as obtained from Ref. [19, is shown in Fig. 4. As mentioned in the introduction, in case of GaP the $E_1$(LO) phonon mode corresponds to coupling of electronic energy states with $\Gamma_{9hh}\rightarrow\Gamma_{7C}$, $\Gamma_{7V}\rightarrow\Gamma_{7C}$ and $\Gamma_{9hh}\rightarrow\Gamma_{8C}$ transitions and the allowed intensity resonance of the $A_1$(LO) mode corresponds to $\Gamma_{7V}\rightarrow\Gamma_{7C}$ symmetry states of the electronic structure [10, 21]. It is to be noted that for GaP band structure, the difference in energy between $\Gamma_{9hh}$ and $\Gamma_{7lh}$ is nearly zero [19]. Thus the resonance of both $E_1$(LO) and $A_1$(LO) modes at 2.67 eV in Fig. 3 can be assigned to coupling of $\Gamma_{7V}\rightarrow\Gamma_{7C}$ states, whereas the other resonance peak of the $E_1$(LO) mode, downshifted by 0.29 eV can be attributed to the coupling of $\Gamma_{9hh}\rightarrow\Gamma_{7C}$ symmetry states in RR measurements. Resonance coupling of $\Gamma_{8C}$ symmetry point is usually small [9, 10, 17] since the larger effective mass of this band compared to that of valence band of $\Gamma_9$ symmetry reduces Raman cross-section of the longitudinal modes coupling these two states. As a consequence, no resonance coupling the valence bands and the $\Gamma_{8C}$ point of the conduction band is observed in our spectra.

The energy states participating in observed electron-phonon coupling are shown by arrows in the schematic band diagram of Fig. 4. The $\Gamma$ point energies that we measure by RR measurements are shown in the figure. In parenthesis we have indicated the same energies at 0K for bulk WZ GaP, as estimated theoretically in Ref. [19. To best of our knowledge, this is the first report demonstrating the expected electron-phonon coupling involving both $\Gamma_{9hh}\rightarrow\Gamma_{7C}$ and $\Gamma_{7V}\rightarrow\Gamma_{7C}$ for the $E_1$(LO) mode and $\Gamma_{7V}\rightarrow\Gamma_{7C}$ states for the $A_1$(LO) mode of WZ electronic structure, as observed in Fig. 3.

In what follows we will call "Raman gap" the energy of the measured $\Gamma_{9hh}\rightarrow\Gamma_{7C}$ resonance. It still does not exclude the conduction band minimum to be of $\Gamma_{8C}$ symmetry, indicating that the optical gap in GaP NW may be smaller than the Raman gap. We find the definition of Raman gap by RR measurements at $E_1$ edge in case of bulk InAs in Ref [25]The difference between optical and Raman gap strictly depends on the phonon involved in the electron-phonon interaction in the system.

Till date all works reported in the literataure, including the discussion above, related to the $\Gamma$ point energies of WZ phase in III-V semiconductor NWs deal with room temperature RR measurements. The experimental results have been compared with the values calculated for 0K. A better comparison with calculated 0K band energies requires temperature dependent RR measurements. Thus, the variation of the resonance energy of $A_1$(LO) mode with temperature was investigated. Raman spectra were collected for four excitation wavelengths over the temperature range between 200 K and 550 K. For different excitation energies, the intensity of $A_1$(LO) mode is maximum at different temperatures (Fig. 5). The variation of resonance energy of the $A_1$(LO) mode coupling $\Gamma_{7V}\rightarrow\Gamma_{7C}$ symmetry states with temperature are shown in Fig. 6. The results are corrected for the incident laser power, the $\omega^4$ law, optical properties of the material and spectral response of the set up for different wavelengths.

The change in band gap with temperature originates from electron–phonon interaction and thermal expansion of the lattice. For III-V semiconductors the variation can be approximated by the empirical relation [26]

$$E_g = E_0 - \frac{\alpha T^2}{\beta + T} \qquad (1)$$

The Varshni parameters $\alpha$ and $\beta$ are constants which strongly depend on material properties, particularly on the crystal structure. $E_0$ is the band gap at 0K. We fitted the data points in Fig. 6 with Eq. 1 and the best fit yields, for the Varshni parameters, $\alpha = 1.0 \times 10^{-3}$ eV.K$^{-1}$ and $\beta = 865$ K. As a reference, the Varshni parameters for the direct energy gap of ZB GaP (obtained by fitting the frequency variation of direct exciton absorption peak with temperature with Eq. 1) are $\alpha = 1.0 \times 10^{-3}$ eV.K$^{-1}$ and $\beta = 825$ K [27], very close to the ones obtained here. This indicates that the band gap of GaP at the $\Gamma$ point has very similar temperature dependence in both WZ and ZB phases.

The most important information conveyed by the fit with Eq. 1 is the value of $E_0 = 2.76 \pm 0.01$ eV. This is an estimate of the A$_1$(LO) resonance (coupling $\Gamma_{7V}$ and $\Gamma_{7C}$) energy at 0K. It is to be recalled that the difference in energy between states with $\Gamma_{7V}$ and $\Gamma_{9hh}$ symmetries of the valence band is ~0.29 eV. Thus, the above analysis yields the value of the Raman gap (gap between $\Gamma_{9hh}$ and $\Gamma_{7C}$) at 0K to be ~2.47 eV, a value more directly comparable to theoretical band energy estimates which are generally calculated at zero temperature.

In addition to band energies, there are other band structure parameters which determine electronic and optical properties of a semiconductor. Several theoretical models, based on self-consistent full-potential linearized-augmented plane wave method within local density approximation, k·p method, are used to estimate the splitting energies and deformation potential in III-V WZ systems [28, 29]. We have estimated the crystal field splitting and spin-orbit splitting energies using the k·p method. Under the quasicubic approximation [29, 10]

$$E_1 - E_{2,3} = \frac{1}{2}\left(\Delta_{cr} + \Delta_{so} \mp \sqrt{(\Delta_{cr} + \Delta_{so})^2 - \frac{8}{3}\Delta_{so}\Delta_{cr}}\right). \quad (2)$$

Here, $\Delta_{cr}$ and $\Delta_{so}$ are crystal field and spin orbit split energy. $E_1$, $E_2$ and $E_3$ are the $\Gamma$ point energies marked in Fig. 4. According to Ref. [19], the value of $E_2$ is very small for GaP. Thus, taking $E_2 \approx 0$ and the experimentally obtained values of $E_1$ and $E_3$, we get $\Delta_{cr}$=0.29 eV and $\Delta_{so}$=0.

Raman gap of WZ GaP at 0K, estimated using density functional theory in the local density approximation (LDA), (which is known to underestimate band gap energies) is 1.81 eV [30]. In a report with full band structure pseudopotential model, the Raman gap is estimated to be 2.88 eV [19]. Our measured zero temperature value of ~2.47 eV falls in between these two estimates, and is not too far from the pseudopotential value. The pseudopotential calculations also yields $\Delta_{cr}$=0.341 eV and $\Delta_{so}$=0.082 eV, not very different from the values obtained in the present work.

In part, the difference can be accounted for considering the empirical expression in Eq. 1, which models the variation of band gap with temperature. This equation is known to have several limitations [31]. This can probably alter our 0K estimate of the Raman gap; however it is unlikely to explain the full difference. Our finding is strongly supported by the fact that for all other III-V semiconductor NWs of WZ phase RR measured gap energy ($\Gamma_{9hh} \rightarrow \Gamma_{7C}$) is lower than the values estimated by De and Pryor in Ref. [19] using pseudopotential calculations. In a recent report the direct band gap (($\Gamma_{9hh} \rightarrow \Gamma_{8C}$) of GaP NWs is measured to be 2.09 eV using photoluminescence measurements [22], while the estimate of Ref. [19] is 2.25 eV. Table II summarizes the experimental measurements on WZ III-V NWs (including the one in the present work) compared to the estimates of De and Pryor (Ref. 19). The measured value of $\Delta_{CR}$ [10] for WZ GaAs NW is also reported to be lower than the estimate in Ref. [19]. The discrepancies between the estimated and measured values of different energy states, crystal field and spin-orbit splitting energies and measured band gap at 0K leave a scope for more rigorous theoretical understanding of the electronic band structure and temperature variation of the band gap in WZ GaP.

## 4. Conclusion

In conclusion, we have measured the Raman spectrum of defect-free WZ GaP NWs and investigated its electronic band structure at Γ points by means of RR measurements. Coupling of electronic energy states through the electron-LO phonon interactions has allowed to determine the energy states of different symmetries in valence and conduction bands. Furthermore, temperature dependent RR measurements allow a comparison of actual band energies with recent theoretical predictions (Ref.19), indicating an overestimation of the Raman gap in latter. This experimental feedback should help to refine theoretical models in order to derive the correct band structure of WZ GaP.

**Acknowledgements**

The work was partly supported by MIUR under PRIN 2009 prot. 2009HS2F7N_003. AR and JKP thank Central Research Facility at IIT Kharagpur for the availability of the Raman spectrometer

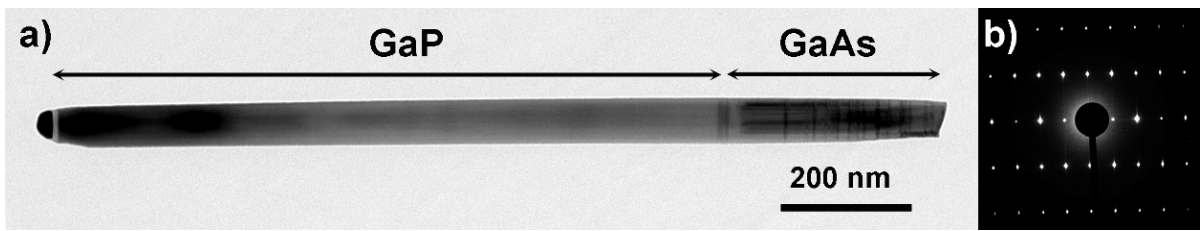

Figure 1. a) Bright field image of a GaP/GaAs NW b) Diffraction pattern collected on the GaP segment

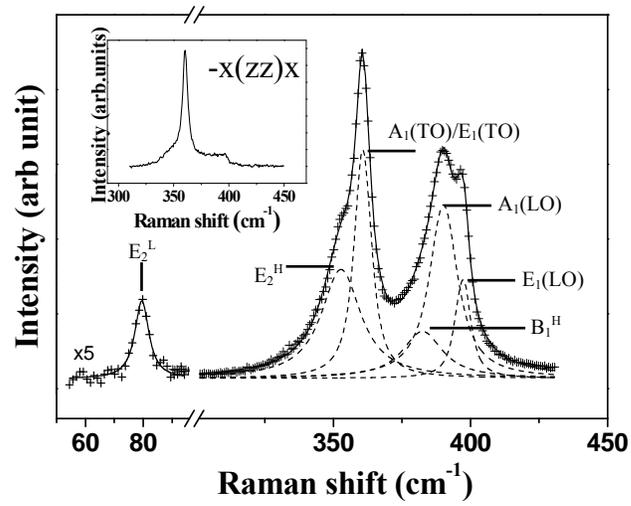

Figure 2. Room temperature Raman spectrum of GaP NWs. Deconvoluted Raman modes are shown by dashed lines. Net fitted spectrum is shown by the solid line. Inset of the figure shows the Raman spectrum of the NW in $x(zz)\bar{x}$ polarization configuration.

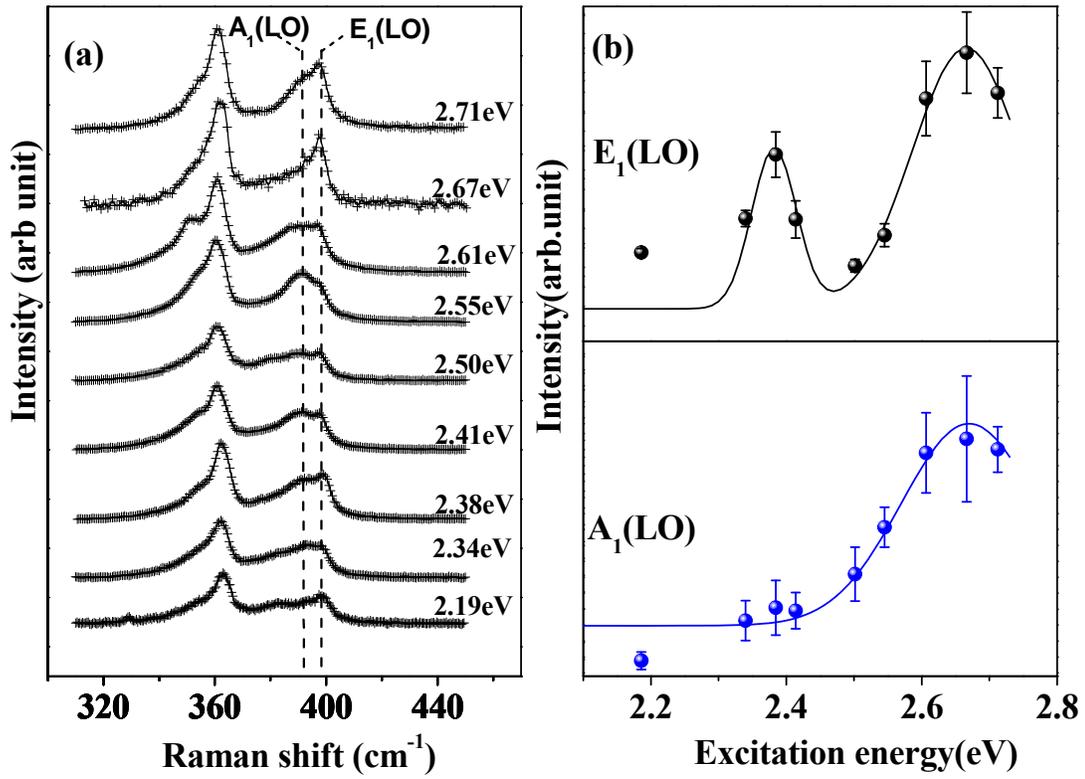

Figure 3. (a) Room temperature Raman spectra of GaP NWs for various excitation energies. (b) Variation of intensity of $A_1$ (LO) and $E_1$(LO) modes with incident excitation energies as obtained from RR measurements.

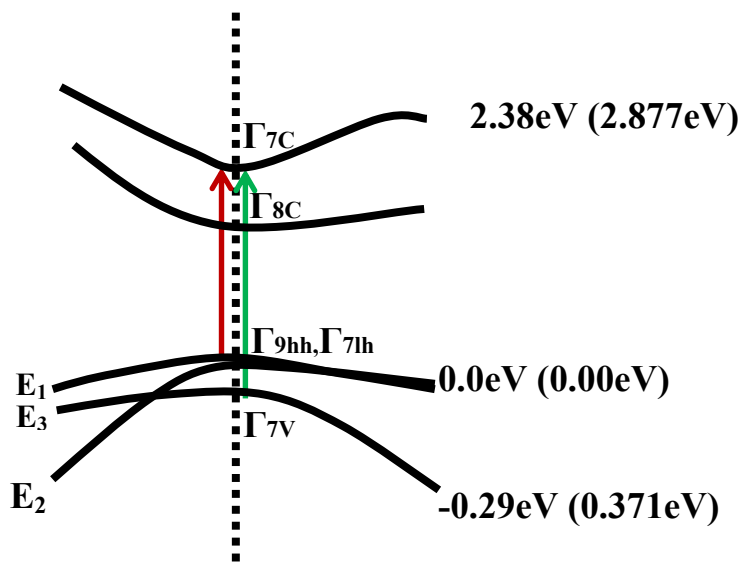

Figure 4. Electronic band structure of GaP in WZ phase at Γ point according to Ref.19 Symmetry and energy of the states, as obtained from RR measurements, at Γ point are marked. The calculated values from reference [19] are shown in the parenthesis. The arrows indicate the transitions observed with RR measurements.

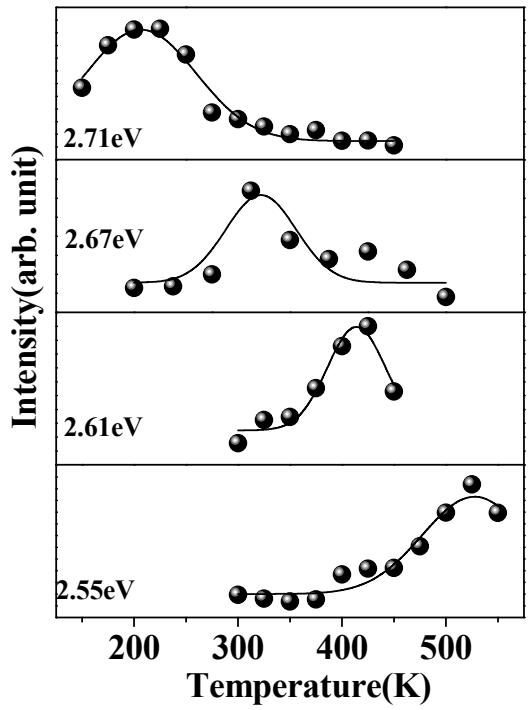

Fig. 5 Intensity of the $A_1$(LO) mode at different temperatures for different excitation energies. The dashed line connects the maxima of each set recorded at different wavelengths. Solid lines are the guide to the eye.

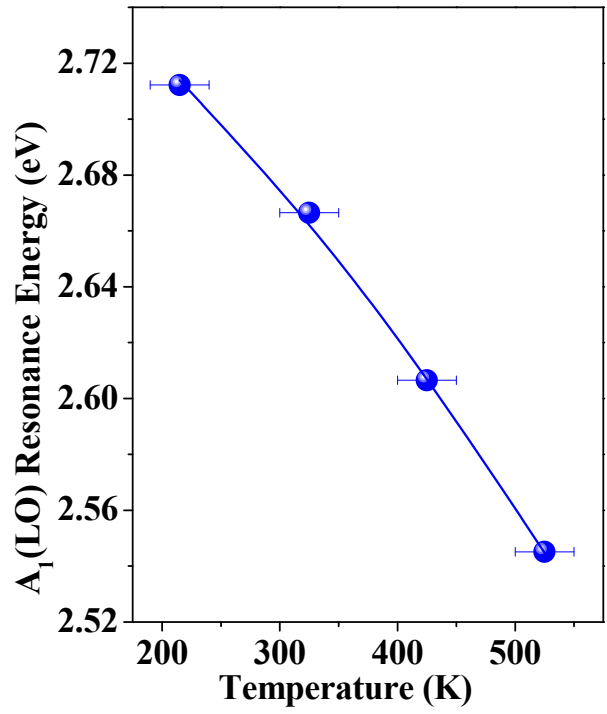

Fig 6. Variation of resonance energy of $A_1(LO)$ mode with temperature, as obtained from Fig. 5. The solid line is the fit to the data point using Eq. 1.

Table I. Expected and measured optical phonon modes of WZ GaP and their polarization configuration.

| Raman mode | Expected | | Measured |
|---|---|---|---|
| | Position (cm$^{-1}$) | configuration | Position (cm$^{-1}$) |
| E$_1$(LO) | 402 | $x(yz)y$ | 397 |
| A$_1$(LO) | <402 | $z(yy)\bar{z}$ | 391 |
| B$_1^H$ | 375 | Silent mode | 383 |
| E$_1$(TO) | 361 | $x(yz)\bar{x}$ $x(yz)y$ | 361 |
| A$_1$(TO) | <361 | $x(yy)\bar{x}, x(zz)\bar{x}$ | 361 $x(zz)\bar{x}$ |
| E$_2^H$ | 355 | $x(yy)\bar{x}, x(yy)\bar{z}$ $z(yx)\bar{z}, z(yy)\bar{z}$ | 353 |
| B$_1^L$ | 215 | Silent mode | – |
| E$_2^L$ | 86 | $x(yy)\bar{x}, x(yy)\bar{z}$ $z(yx)\bar{z}, z(yy)\bar{z}$ | 80 |

Table II. A comparison between measured energy gap for III-V NWs of WZ phase as obtained from RR and photoluminescence (PL) and the theoretical estimates of Ref. [19]. Superscript are the references for experimental measurements. In bold the results from the present work.

| System | symmetry states involved | Measured value(eV) | Estimated 0K value (eV)[19] |
|---|---|---|---|
| GaAs NW | ($\Gamma_{9hh} \rightarrow \Gamma_{7C}$) | *1.56[9] *1.462[7] | 1.588 |
|  | ($\Gamma_{7lh} \rightarrow \Gamma_{7C}$) | *1.561[7] | 1.708 |
| AlAs NW | ($\Gamma_{9hh} \rightarrow \Gamma_{7C}$) | *1.96[17] | 3.153 |
|  | ($\Gamma_{9hh} \rightarrow \Gamma_{8C}$) |  | 1.971 |
| GaP NW | **($\Gamma_{9hh} \rightarrow \Gamma_{7C}$)** | **2.47**† | 2.877 |
|  | **($\Gamma_{7V} \rightarrow \Gamma_{7C}$)** | **2.76**† | 3.250 |
| GaP NW(PL) | ($\Gamma_{9hh} \rightarrow \Gamma_{8C}$) | 2.09[22] | 2.25 |

*RT measurements    †present work, 0K extrapolations